\documentclass{iopart}

\usepackage{graphicx,amsfonts}
\usepackage{bm}

\newcommand{\pot}{^{40}}
\newcommand{\hfsplit}{ \Delta_{ \mathrm{HF} }   }
\newcommand{\ket}[1]{\displaystyle{|#1\rangle}}
\newcommand{\bra}[1]{\displaystyle{\langle #1|}}
\newcommand{\proj}[2]{\displaystyle{| #1 \rangle \langle #2|}}

 \def\ee{\mathord{\rm e}}
 
 \def\ii{\mathord{\rm i}}

\begin{document}

\title[A Quantum Simulator for Relativistic Field Theories and Topological Insulators]{An Optical-Lattice-Based Quantum Simulator For Relativistic Field Theories and Topological Insulators}

\author{Leonardo Mazza}
\address{Max-Planck-Institut f\"ur Quantenoptik, D-85748 Garching, Germany}

\author{Alejandro Bermudez}
\address{Institut f\"ur Theoretische Physik, Universit\"at Ulm, D-89069 Ulm, Germany}

\author{Nathan Goldman}
\address{Center for Nonlinear Phenomena and Complex Systems - Universit\'e Libre de Bruxelles, B-1050 Brussels, Belgium}

\author{Matteo Rizzi}
\address{Max-Planck-Institut f\"ur Quantenoptik, D-85748 Garching, Germany}

\author{Miguel Angel Martin-Delgado}
\address{Departamento de F\'isica Te\'orica I, Universidad Complutense, E-28040 Madrid, Spain}

\author{Maciej Lewenstein}
\address{ICFO - Institut de Ci\`encies Fot\`oniques, E-08860 Castelldefels (Barcelona), Spain}
\address{ICREA - Instituci\'o Catalana de Recerca i Estudis Avan\c{c}ats, E-08010 Barcelona, Spain}

\begin{abstract}
We present a proposal for a versatile cold-atom-based quantum simulator of relativistic fermionic theories and topological insulators in  arbitrary dimensions. 
The setup consists of a spin-independent optical lattice that traps a collection of hyperfine states of the same alkaline atom, to which the different degrees of freedom of the field theory to be simulated are then mapped. 
We show that the combination of bi-chromatic optical lattices with Raman transitions can allow the engineering of a spin-dependent tunneling of the atoms between neighboring lattice sites.
These assisted-hopping processes can be employed for the quantum simulation of various interesting models, ranging from non-interacting relativistic fermionic theories to topological insulators.  We present a toolbox for the realization of different types of relativistic lattice fermions, which can then be exploited to synthesize the majority of phases in the periodic table of topological insulators.
\end{abstract}

\pacs{37.10.Jk, 11.15.Ha, 72.20.-i}

\submitto{NJP}

\maketitle

\section{Introduction \label{sec:introduction}}

In a seminal paper published in 1982~\cite{Feynman}, R. P. Feynman discussed in great detail the problems connected with the numerical simulation of  quantum  systems. He envisaged a possible solution, the so-called \emph{universal quantum simulator}, a quantum-mechanical version of the usual simulators and computers currently exploited in many applications of the ``classical'' world.
If realized, such a device would be able to tackle many-body problems with local interactions by using the quantum properties of nature itself~\cite{Lloyd}. 
Interestingly, even without the advent of a fully universal quantum computer, the construction of small dedicated devices, also known as \textit{purpose-based quantum simulators}, would already be of significant importance for  our understanding of quantum physics.
The basic idea is to engineer the Hamiltonian of the quantum model of interest in a highly-controllable quantum system, and to retrieve all the desired information with a measurement of its properties.
Many research fields would eventually benefit from such devices, such as two- and three-dimensional many-body physics, non-equilibrium dynamics, or lattice gauge theories~\cite{Nori}. 

Recently, the scientific community is considering ultra-cold atoms 
as one of the most promising candidates for the realization of a wide variety of dedicated quantum simulations~\cite{BDZ, Lewenstein}. 
Indeed, these  gases are genuine quantum systems where the available experimental techniques offer an impressive degree of control together with high-fidelity measurements, thus combining two fundamental requirements for a quantum simulator.
Among the most recent experimental achievements, we would like to mention the observation of Anderson localization in disordered Bose-Einstein condensates (BEC)~\cite{Aspect, anderson_atoms}, the research on itinerant ferromagnetism with cold fermions~\cite{ItinFerro}, or  the reconstruction of the equation of state of fermionic matter in extreme conditions, such as in neutron stars~\cite{Salomon}. 

An important drawback in the applicability of cold atoms as quantum simulators is the difficulty of coupling their spatial degrees of freedom to external magnetic fields. 
This prevents a direct simulation of the quantum Hall physics~\cite{QHE}, the controlled observation of whose extraordinary phenomenology would shed new light on quantum many-body theory. 
One  way to overcome this problem is to dress the system with ingenious laser schemes, which mimic the effect of an external magnetic field, and thus allow the neutral  atoms to behave as effectively charged particles~\cite{Gauge}. This approach led recently to the realization of neutral BECs coupled to external effective magnetic and electric fields~\cite{Spielman1, Spielman2}, or even with an effective spin-orbit coupling~\cite{Spielman3}.
More generally, the scientific community has now realized that even in presence of an optical lattice, dressing cold gases with suitable optical and microwave transitions could push the experiments beyond the standard superfluid - Mott insulator transition, and significantly widen the spectrum of the models that are currently being simulated~\cite{ConfKI}. The possible applications of such optical-lattice-based quantum simulations are numerous and diverse, ranging from the realization of Abelian and non-Abelian static gauge fields~\cite{Gauge, JakschZoller, GerbierDalibard, NonAbelian, GoldWilson} to that of quantum Hall states~\cite{SorDemLuk, Hafezi, Palmer, MollCooper, Umucalilar}; from the study of the anomalous quantum Hall effect~\cite{GoldAnom, LiuLiu} to the quantum spin Hall effect~\cite{SpielmanTI, Stanescu}; from  three-dimensional topological insulators~\cite{MazzaRizziPRL}, to flat-band physics with a non-trivial topological order~\cite{Sun},  or non-Abelian anyons~\cite{Burrello}. Recently, a big effort has also been put in designing schemes where the exotic effects associated to relativistic quasiparticles, such as the  Klein tunneling and the \textit{Zitterbewegung}, arise in a controlled table-top  experiment~\cite{Hou, Lim, Lepori, PachosIgnacio, Celi, dirac_fermions, Witthaut, Schuetzhold}.

In this article, we elaborate on the idea of using a spin-independent bi-chromatic optical lattice dressed with suitable Raman transitions to simulate interesting non-interacting field theories of lattice fermions. We present a concrete proposal to create a three-dimensional optical lattice that traps a multi-species atomic gas, and to tailor arbitrary spin-dependent hopping operators.
We have already shown how this setup could break the SU(2) invariance of the hopping rates for spin-1 atoms in spin-independent lattices, and how the simulation of systems subjected to three-body repulsion could benefit from it~\cite{MazzaRizziPRA}. Here, we extend this idea further, and show that the same setup  allows for the realization of hopping operators which modify the atomic hyperfine state. 
Combining this trapping scheme with Fermi gases, we show that this platform would open a new route towards the simulation of high-energy physics and topological insulators.

This paper is organized as follows: in section~\ref{sec:Setup}, we describe qualitatively the idea of using an optical superlattice  to realize a general hopping operator for a multi-species cold gas of alkalis. Further analysis and technical details are given in section~\ref{sec:NonTrivial}, where we also present  some numerical results that support the possibility of controlling a spin-flipping tunneling in this platform. The reader not interested in these technical details may skip this content without prejudicing the comprehension of the following sections. Some final remarks on the proposal are presented in section~\ref{sec:critical}. In section~\ref{sec:proposals}, we discuss  the possible applications of the described scheme, focusing on relativistic theories and topological insulators, and trying to give a list of the most interesting phenomena which could be explored. Finally, we present our conclusions in section~\ref{sec:conclusion}.

\section{The Setup and The Idea \label{sec:Setup}}

\begin{figure}[t]
\begin{center}
\includegraphics[width=\textwidth]{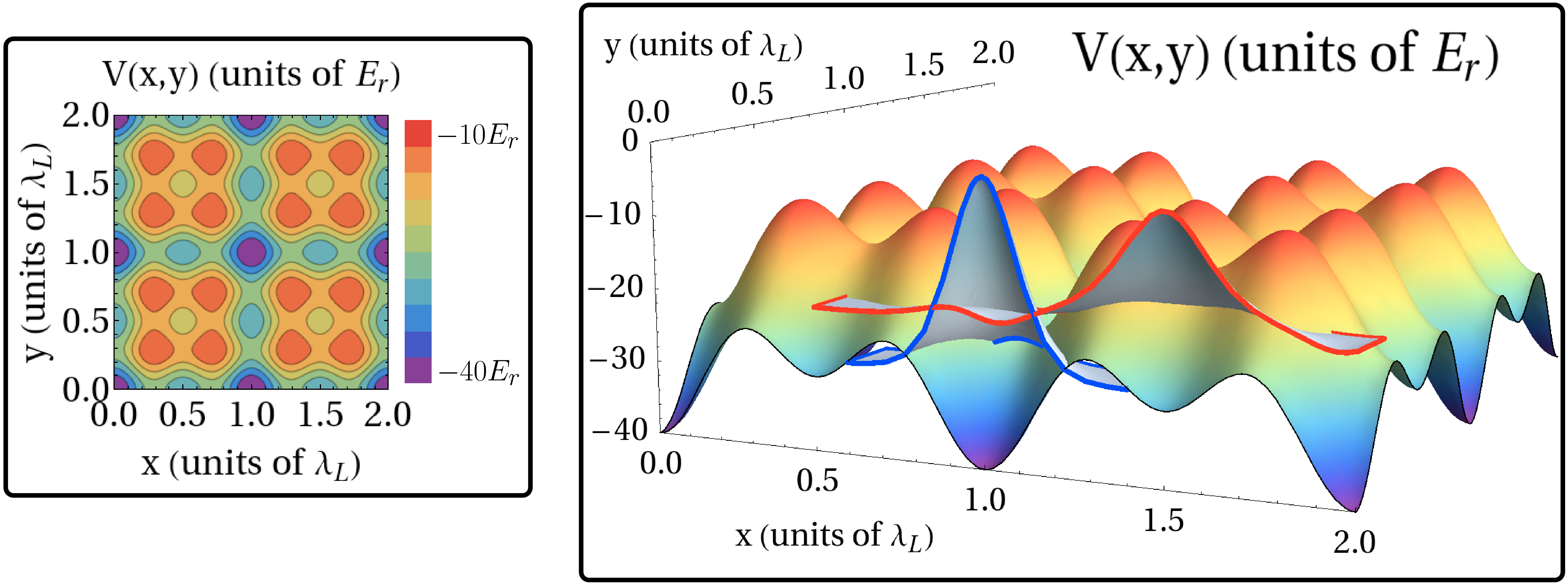}
\caption{Optical superlattice potential of equation~(\ref{eq:potential}) in the
two-dimensional case, with parameters $V_0=10E_r$ and $\xi=1$. Left: the
potential is characterized by a square geometry of main minima; in the middle of
each link an intermediate minimum is also present. Right: if the lattice is deep
enough, the spectrum of the system features two energy bands whose Wannier
functions are localized in the main minima and in the secondary minima, as
plotted in the figure.}
\label{fig:Superlattice2D}
\end{center}
\end{figure}

We consider the following atomic three-dimensional optical potential
\begin{equation}
V(\mathbf x) = - V_0 \sum_{j \in\{1,2,3\}} \left[ \cos^2 (q x_j) + \xi \cos^2 (2 q x_j) \right],
\label{eq:potential}
\end{equation}
where  $\mathbf x = (x_1, x_2, x_3)$, $q = 2 \pi / \lambda_{L}$ ($\lambda_L$ is the wavelength of the laser), and where $V_0, \xi >0$ represent the potential amplitudes.
The low-energy structure of this potential is a cubic array of main minima separated by ``secondary'' minima
located in the middle of each lattice link (see figure~\ref{fig:Superlattice2D}). We note that additional higher-order minima are also present, but will not play any role in the phenomena discussed in this article.
Due to the specific form of the potential in equation~(\ref{eq:potential}), the Hamiltonian can be divided into three independent terms, each one depending on one of the three couples of conjugate operators, $\{x_i, p_i \}_{i\in 1,2,3}$. Consequently, the Bloch functions of the $n$-th band with energy $E_n(\mathbf p)$, can be written as $\psi_{n,\mathbf p} (\mathbf x) = \prod_j \psi_{n,p_j}(x_j)$. 
In order to discuss the effects occurring on the scale of one lattice site, Wannier functions can be introduced for each band
\begin{equation*}
w_{n,\mathbf R} (\mathbf x) = \frac 1{V} \int e^{-i \mathbf{R \cdot p}}\psi_{n,\mathbf p} (\mathbf x) d\mathbf p=
w_{n,R_1} (x_1) \, w_{n,R_2} (x_2) \, w_{n,R_3} (x_3).
\label{eq:wannier_factor}
\end{equation*}
Like Bloch functions, Wannier functions belonging to different bands form an orthonormal basis, and one can thus expand the Hamiltonian in such a basis. Since the Wannier functions are not eigenstates of the Hamiltonian, this  expansion  leads to a Hubbard model  describing the tunneling of atoms between neighboring sites, together with a local on-site interaction coming from the scattering of the cold gas~\cite{CiracOL}.  

\begin{figure}[t]
\begin{center}
\includegraphics[width=\textwidth]{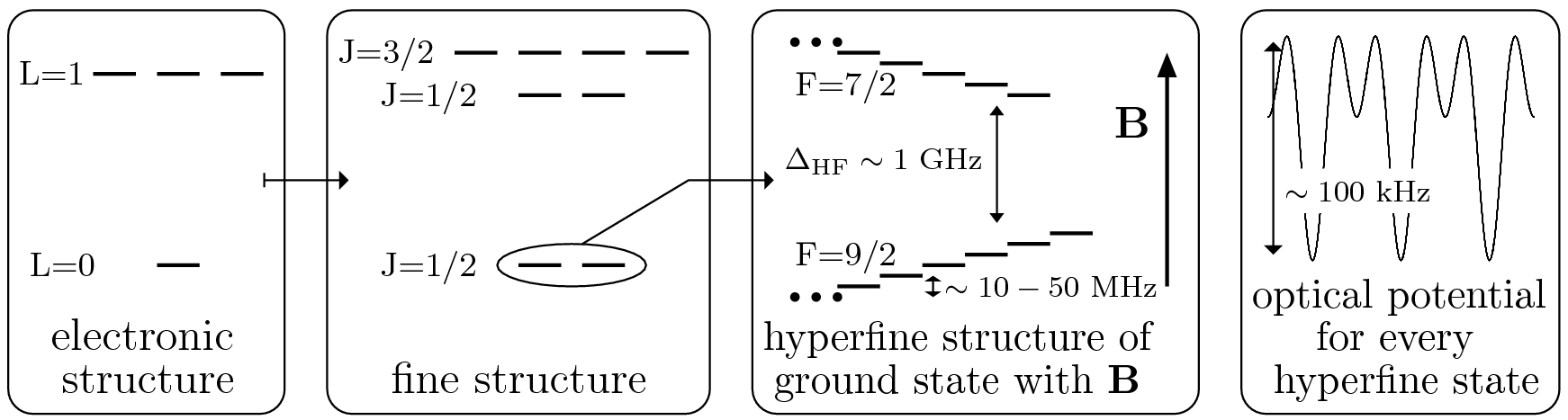}
\caption{Sketch of the atomic structure of $\pot$K: from the electronic structure ($\mathbf L$ is the electronic angular momentum) to the fine structure ($\mathbf J = \mathbf L + \mathbf S$, where $\mathbf S$ is the electronic spin) to the hyperfine structure ($\mathbf F = \mathbf J + \mathbf I$, where $\mathbf I$ is the nuclear spin). The latter is drawn in the specific case of an external magnetic field present. The last box shows the optical spin-independent potential which traps equally all the hyperfine levels. }
\label{fig:AtomiclevelsK40}
\end{center}
\end{figure}

This setup can be used for the simulation of a lattice field theory, where the field operators are identified with the atomic creation-annihilation operators in the Wannier basis of the lowest energy band (i.e. the states localized in the main minima of the lattice). Conversely, higher energy bands  provide auxiliary levels that shall be used as a resource to tailor the tunneling processes.
The main result of this article is the claim that a complicated though not unfeasible combination of current technologies leads us to the realization of the following Hamiltonian
\begin{equation}
H_{\mathrm{sys}}=
\sum_{\mathbf{r}\bm{\nu}}\sum_{\tau\tau'}
t_{\nu}c^{\dagger}_{\mathbf{r}+\bm{\nu}\tau'}[U_{\bm{\nu}}]_{\tau'\tau}c_{\mathbf{r}\tau}+\sum_{\mathbf{r}}\Omega c^{\dagger}_{\mathbf{r}\tau'}[\Lambda]_{\tau'\tau}c_{\mathbf{r}\tau}+\mathrm{H.c.}
\label{eq:claim}
\end{equation}
Here, we are considering a multi-species fermionic scenario with many hyperfine levels of the same atom: $c^{\dagger}_{\mathbf{r}\tau}$ $(c_{\mathbf{r}\tau})$ creates (annihilates) a fermion with hyperfine spin $\tau$ localized in the main minima of the superlattice at  $\mathbf{r}=m_1\mathbf{a}_1 + m_2 \mathbf{a}_2 + m_3 \mathbf{a}_3$, where $m_j\in\{1...L_j\}$, $L_j$ stands for the number of lattice sites along the $x_j$ axis, and $\mathbf a_j$ is the lattice spacing in the $j$-th direction. 
The parameter $t_{\nu}$ stands for the strength of the laser-assisted tunneling in the $\hat{\bm{\nu}}$ direction, with with $\bm{\nu}\in\{\mathbf{a}_1,\mathbf{a}_2,\mathbf{a}_3\}$,  which shall be described below. The operators $U_{\bm{\nu}}$ describe the tunneling from $\mathbf r$ to $\mathbf r + \bm{\nu}$, and are a common feature in lattice gauge theories. We have also included an on-site Raman term $\Lambda$, of strength $\Omega$, that induces a certain transition between the hyperfine states. Note that we use Gaussian units and $\hbar=1$.
The claimed possibility of engineering a wide range of hopping operators $U_{\bm{\nu}}$, together with the state-of the-art control of the atomic interaction, makes already our system 
a versatile quantum simulator of  lattice field theories. In this manuscript, we focus on
 non-interacting theories, which can be realized either with dilute systems, or by employing  Feshbach resonances to lower the interaction strength (see e.g.~\cite{Aspect,anderson_atoms}). 
We stress that interesting phenomena can also be  observed in non-interacting gases when additional ingredients are introduced in their dynamics, such as disorder or the assisted-hopping processes discussed in this article.

Let us note that the control of the homogeneous tunneling  for a single-species atomic gas is straightforward, and would not even require the superlattice ($\xi = 0$)~\cite{Koehl}. Moving to a many-species case, one runs into the problem that a general hopping operator also entails  terms flipping the atomic hyperfine spin (simply referred as \emph{spin} in the following), which are not easily engineered. Here, we propose to realize such couplings by combining Raman transfers and a bi-chromatic superlattice ($\xi \neq 0$ in equation~(\ref{eq:potential})). The proposal can be applied to all the alkalis notwithstanding their bosonic or fermionic nature. In the following, however, we shall focus in the fermionic scenario, which is best explained with the following practical example.

\begin{figure}[t]
\includegraphics[width=\textwidth]{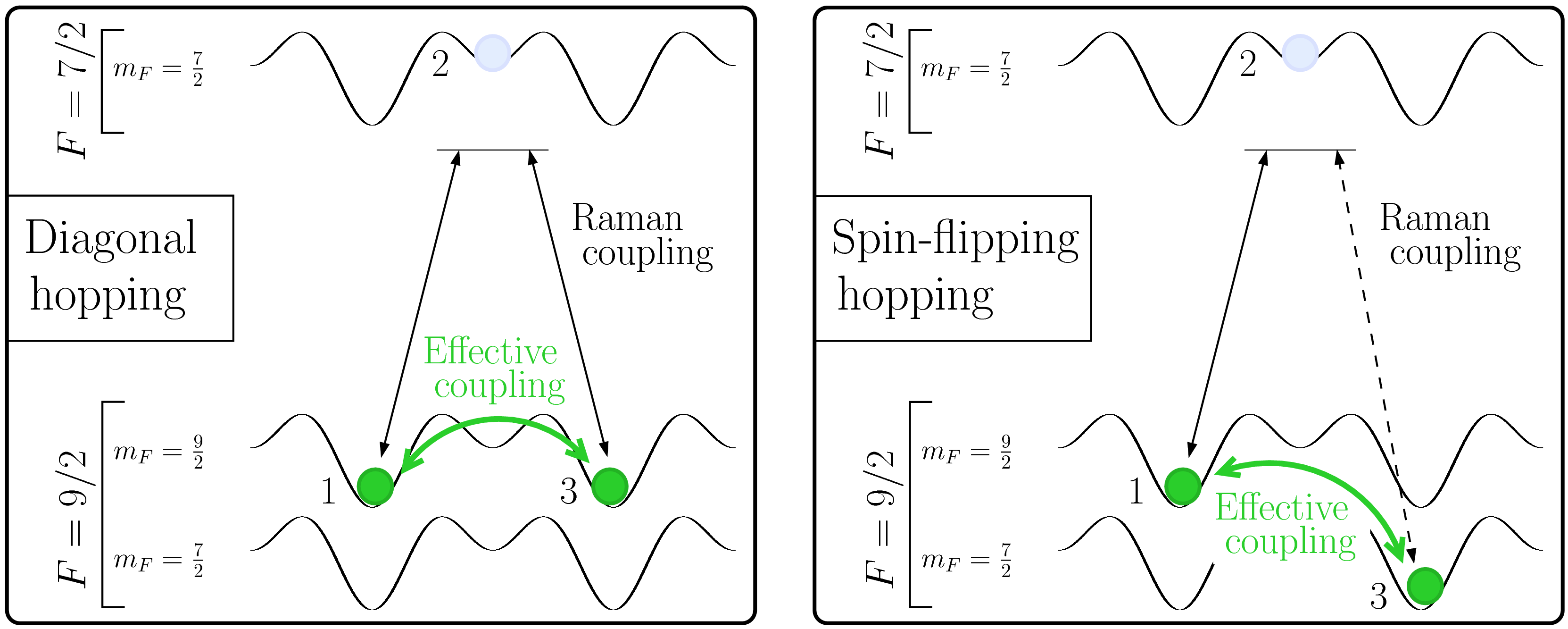}
\caption{Sketch of a laser-assisted tunneling induced in the presence of a superlattice.
Two physical hyperfine states belonging to the $F=9/2$ manifold are connected via Raman couplings with the intermediate level of an auxiliary state belonging to the $F=7/2$ manifold.
If the coupling is detuned enough, the $F=7/2$ level can be adiabatically eliminated: no population is left there and an effective coupling is engineered between neighboring sites. Left: scheme for a spin-preserving (i.e. diagonal) hopping. Right: scheme for a spin-flipping hopping.}
\label{fig:sketch}
\end{figure}

Let us consider an ultra-cold cloud of non-interacting $\pot$K atoms in the presence of a magnetic field of intensity $B$. Such field  
lifts the spin degeneracy within the two atomic hyperfine 
manifolds of the ground state, $F = 9/2$ and $F=7/2$,
according to the following relations (see also figure~\ref{fig:AtomiclevelsK40}):
\begin{equation}
E_{9/2, m_F} =    +  g_F \, \mu_B \, B \; m_F \qquad
E_{7/2, m_F} =  \Delta_{{\rm HF}}   -  g_F \, \mu_B \, B \; m_F 
\label{eq:hyperfinesplitting}
\end{equation}
where $m_F$ is the projection of the hyperfine spin along the quantization axis defined by  the magnetic field, $\mu_B$ is the Bohr magneton, $g_F$ is the hyperfine Land\'e Factor, and $\Delta_{{\rm HF}}$ stands for the hyperfine splitting. These hyperfine levels are all trapped into the same spin-independent optical potential~(\ref{eq:potential}).
Depending on the lattice theory we want to simulate, we select a subset of these hyperfine levels described theoretically by creation-annihilation operators in the lattice sites. We then identify such fields with the components of the  lattice field theory to be simulated. This leads us to divide the hyperfine levels into two subsets: the subset of ``physically meaningful'' states, which belong to the hyperfine manifold $F=9/2$, and the usually larger subset of auxiliary levels that shall be used to assist the tunneling and create  the desired hopping operator.

\begin{table}[t]
\caption{Numerical values of a possible three-dimensional optical bi-chromatic superlattice~(\ref{eq:potential})
for $\pot$K used in section~\ref{sec:NonTrivial} for numerical simulations.
We characterize the properties of the two energy bands, which exhibit Wannier functions localized respectively in the main and secondary minima, by 
listing
the energy expectation value $\langle E_i \rangle$ of the most localized Wannier function and the bandwidth $\Delta E_i$.
Finally, we argue that atoms trapped in optical lattices show a hierarchy of typical energies which can be actively exploited for engineering non-trivial hopping operators.
\label{table:K40}
\label{table:superlattice}}
\begin{indented}
\item
\begin{tabular}{cc|cc}
\br
$\lambda_{L}$ & $\sim 738$ nm & $E_{r} = \frac{h^2}{2 m \lambda_{L}}$ & $9.17$ kHz \\
$ V_0 $ & $10$ $E_r$ & $\xi$ & 1 \\
\mr
$\langle E_{1} \rangle$ & $-13.909 * 3$ $E_r$ & $\langle E_{2} \rangle$ & $-13.909*2 -6.364$ $E_r$ \\
$ \Delta E_{1} $ & $0.024$ $E_r$ & $ \Delta E_{2} $ & $0.995$ $E_r$ \\
$ \Delta E_{1} $ & $216.7$ Hz & $ \Delta E_{2} $ & $9120.1$ Hz \\
\mr
$\hfsplit$ & $1.286$ GHz & $g_F \, \mu_B$ & $0.22*1.34$ MHz/G \\
$\langle E_{2} - E_{1} \rangle$ & $69.160$ kHz & Staggering & $10$ kHz \\
\br
\end{tabular}
\end{indented}
\end{table}

Regarding the hopping operator in equation~(\ref{eq:claim}), we address each of its matrix elements
$[U_{\bm{\nu}}]_{\tau'\tau}$
 separately. Given a matrix element (i.e.
once we have identified the initial and final hyperfine levels to be connected by the assisted tunneling), 
we choose an auxiliary level belonging to the hyperfine manifold $F=7/2$   trapped in the middle of the link. These levels  provide intermediate ``bus'' states that shall be used as a resource to assist the tunneling as follows. 
The couplings between the atoms in the main sites, $\mathbf R_1$, and the ``bus'' states, $\mathbf R_2$,  are realized via optical two-photon Raman processes transferring a net momentum $\mathbf q_t$. They have a mathematical expression  proportional to the overlap integral of the initial and final Wannier functions:~$
\int w^*_{n_2,\mathbf R_2}(\mathbf x) e^{i \mathbf q_t \cdot \mathbf x} w^{\phantom{*}}_{n_1,\mathbf R_1}(\mathbf x) d \mathbf x
$.
This integral is not zero because of the term $e^{i\mathbf{q}_t\mathbf{\cdot x}}$, which is of course relevant only if 
$2\pi/|\mathbf q_t|$ is of the order of the lattice spacing. Since this regime cannot be achieved with microwave transitions, one is motivated to employ two-photon Raman transitions.
Interestingly enough, it is possible to  eliminate adiabatically the intermediate level and obtain an effective four-photon coupling between neighboring sites (see figure~\ref{fig:sketch}). We stress that different matrix elements can be engineered at the same time thanks
to the magnetic-field splitting of the hyperfine levels~(\ref{eq:hyperfinesplitting}):
the involved atomic transitions become non-degenerate and can be 
individually addressed with different lasers. Furthermore, the use of coherent laser light for the Raman transitions entails the additional advantage of being able to deal with complex phases, and thus to realize complex gauge structures at will.
The realization of the non-diagonal matrix elements requires the lattice to be slightly staggered, a technique discussed also in reference~\cite{GerbierDalibard}.
Summarizing, this proposal tries to exploit a hierarchy of energies characterizing atomic gases in optical lattices in order to assist the tunneling between neighboring sites with controlled adiabatic eliminations (see table~\ref{table:K40}). 

The on-site spin-flipping $\Lambda$ in equation~(\ref{eq:claim}) can be performed with  standard technology based on microwave transitions, or  Raman transitions carrying negligible momentum. Furthermore, these terms can also be  exploited  to correct spurious on-site couplings which may be induced by the laser scheme.
Unfortunately, we note that there is no selection rule relying on the polarization properties of the light and the hyperfine moment of the atoms, which can be used to realize the different  on-site and nearest-neighbors spin-flipping processes. The superimposed magnetic field cannot be aligned at the same time with the propagation vector of all the three lasers, aligned along the three Cartesian axes, which would be the case in which circularly polarized light could be exploited to induce controlled transitions. Conversely, we shall rely on the different Zeeman-shifted energies to selectively address the different couplings between the internal states.

\section{Realization of Spin-Dependent Hopping Operators \label{sec:NonTrivial}}

In this technical section, we theoretically and numerically 
confirm the qualitative scheme presented above.
We  study two simple but important cases: 
the realization of diagonal and non-diagonal hopping operators for a two-species atomic gas. 
These can be considered as the main building blocks needed to realize any tunneling operator even in situations with more than two atomic species.

\subsection{Coupling Between Different Hyperfine Manifolds}

The most fundamental ingredient of this proposal is the possibility of using Raman processes to induce controlled atomic transitions
between different hyperfine states of the electronic ground state $L=0$ 
($L$ is the total electronic angular momentum).
These transitions are realized with two lasers via adiabatic elimination of
the electronically excited manifold $L=1$. In the following, we address the atomic levels as $\ket{L, \alpha, k}$, with $\alpha$ labeling the hyperfine degrees of freedom  (see also figure~\ref{fig:AtomiclevelsK40} for some insights on the internal structure of $\pot$K), and $k$ the quantum numbers of the center-of-mass wavefunction (in our case, the Wannier functions of the optical potential).
As discussed in~\cite{MazzaRizziPRA}, the induced Raman coupling between the state $\ket{0, \alpha, k}$ and $\ket{0, \alpha', k'}$ can be written as follows:
\begin{equation}
\tilde{\Omega}_{\alpha' k'; \alpha k} (t)
= S_{k'k} \; \Omega_{\alpha' \alpha} \; e^{- i \omega t} \phantom{ciao ciao ciao ciao ciao ciao ciao c}
\label{eq:shortRaman}
\end{equation}
This expression clearly factorizes the following contributions:
\begin{itemize}
\item the time-dependence of the effective coupling and its effective frequency, which is the difference between the frequencies of the two lasers $\omega = \omega_1 - \omega_2$;
\item the dependence on the center-of-mass degrees of freedom, $S_{k'k} = \bra{k'} e^{-i (\mathbf p_2 - \mathbf p_1) \cdot \mathbf x} \ket{k}$, where $\mathbf p_1$ and $\mathbf p_2$ are the momenta of the two lasers;
\item the dependence on the initial and final internal states and on the polarization properties of light, $\Omega_{\alpha' \alpha}$, which is a function of the dipole matrix elements between the initial (final) state and the excited levels.
\end{itemize}
Next, we specify~(\ref{eq:shortRaman}) to the superlattice setup of section~\ref{sec:Setup}, i.e. we will consider Raman transitions in presence of lattices characterized by a Wannier function trapped in the middle of each link.

\subsection{Developing an Effective ``6-Level Model''}

\begin{figure}[t]
\begin{center}
\includegraphics[width=0.85\columnwidth]{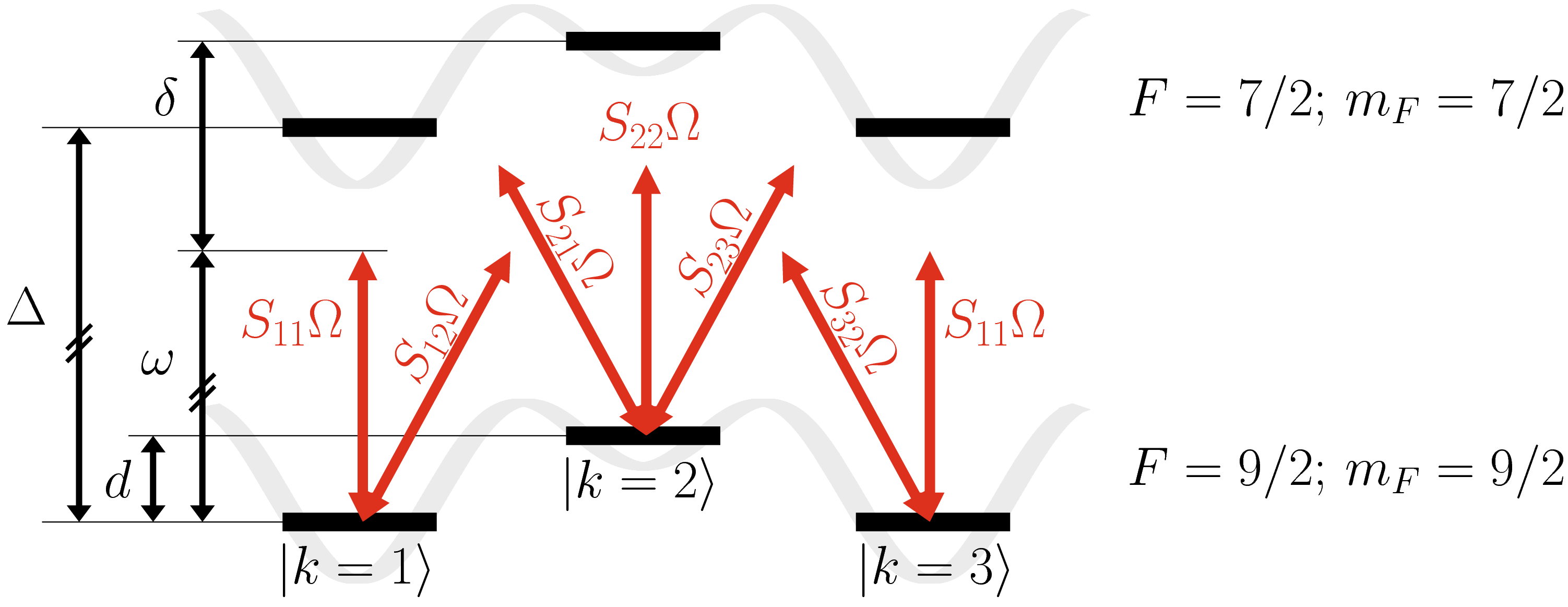}
\caption{The ``6-level model'' used to model the spin-preserving (diagonal) hopping of $F=9/2, m_F=9/2$. The auxiliary state $F=7/2, m_F = 7/2$ has been chosen. The center-of-mass quantum number is $k$. Energies are not in scale; the orders of magnitude of the parameters are the following: $d \sim 10\div 100$ kHz, $\delta \sim 100 \div 300$ kHz and $\Delta \sim 1\div 10$ GHz. We propose to adiabatically eliminate the upper manifold and to study the dynamics of the lowest one with an effective Hamiltonian $H_{pert}$~(\ref{eq:pert}).}
\label{fig:link}
\end{center}
\end{figure}

Let us address the simulation a theory characterized by two-component fields.
Following the discussion of section~\ref{sec:Setup}, we take two states of the $F=9/2$ manifold of $\pot$K, for instance $\ket{ 9/2; \, m_F = 7/2}$ and $\ket{9/2;\, m_F = 9/2}$, and map them into the theory to be simulated.
Here and in the following subsections, we discuss the laser-assisted hopping in the diagonal case ($m_F$ preserved while hopping) and non-diagonal case ($m_F$ flipped while hopping).

For the diagonal case, 
we develop the ``6-level model'' depicted in figure~\ref{fig:link}.
We consider one physically meaningful state, say $\ket{F=9/2, m_F=9/2}$, and one auxiliary state, say $\ket{F=7/2, m_F=7/2}$. Moreover,  we consider  different
Wannier states for each of them, two localized in main sites ($k=1$ and $3$) and one in the intermediate link ($k=2$).
The model includes the effects of undesired couplings and additional levels, and
its limitations, together with the approximations on which it relies,
will be discussed at the end of the paragraph.
We can identify the states with the short notation $\ket{F,k}$ rather than with the longer previous one $\ket{0 \, \alpha \, k}$.
Below, we give an analytical estimate of the population transfer rate, whereas in the next subsections we present the numerical time-evolution for physically interesting cases.

The model is parametrized by six relevant couplings between the different Wannier functions $S_{k'k}$ (see figure~\ref{fig:link}),
whose properties are listed  below. We exploit the existence of theorems which assure the possibility, in our case, of considering three real and exponentially localized Wannier functions $w_j(\mathbf x), j \in \{1,2,3\}$~\cite{Kohn}. We write the parameters $S_{k'k}$ factorizing out the space dependence of the coupling $e^{i \mathbf q_t \cdot \mathbf x_j}$, where $\mathbf x_j$ is the position of the point around which the Wannier function $w_j(\mathbf x)$ is localized,
\begin{eqnarray}
S_{k'k} = e^{i \mathbf q_t \cdot \mathbf x_k} \int 
w_{k'}^* (\mathbf{x-x}_{k'}+\mathbf x_{k}) e^{i \mathbf q_t \cdot \mathbf x} w_k(\mathbf{x})
d \mathbf x; \label{eq:S} \\
S_{1,1} = S_{3,3} \neq S_{2,2}; 
\qquad
S_{1,3}, \; S_{3,1} \sim 0.
\end{eqnarray}
The parameters $S_{1,1}$ and $S_{2,2}$ describe two on-site couplings, whereas $S_{1,2}$ is the coupling between a main site and an intermediately trapped state (see figure~\ref{fig:link}).
The last relation states that couplings between neighboring main sites are negligible. 
The relation between the other four overlap factors depends on the particular experimental situation. In this case, we are interested in the simplest scenario where a single Raman transition induces all these couplings, which leads us to  $$S_{1,2} = S_{2,1} = e^{ 2 i \mathbf q_t \cdot \mathbf x_1} e^{i \mathbf q_t \cdot ( \mathbf x_2 - \mathbf x_1)} S_{3,2}^* = e^{ 2 i \mathbf q_t \cdot \mathbf x_1} e^{i \mathbf q_t \cdot ( \mathbf x_2 - \mathbf x_1)} S_{2,3}^*.$$ 
In order to make this scheme simpler, we assume $\mathbf q_t = 2 \mathbf q_L$, and thus $e^{i \mathbf q_t \cdot ( \mathbf x_2 - \mathbf x_1)} = 1$. As we will argue below, transferring a momentum which does not fulfill this requirement is not a problem since the resulting phase can be gauged away. The phase $2 \mathbf q_t \cdot \mathbf x_1$ can also be put to zero for the moment, since its role only becomes important when one needs to give a phase to different matrix elements.
In the following, we will also consider situations where the coupling between the lattice sites $2$ and $3$ could be induced by lasers propagating in the opposite direction, where $S_{1,2} = S_{2,1} =  S_{3,2} = S_{2,3}$. Taking these considerations into account, the Hamiltonian reads as follows (see figure~\ref{fig:link} for the definitions of $\delta$, $\Delta$ and $\omega$):
\begin{eqnarray}
H &=& d \; \proj{9/2,2}{9/2,2} + (\Delta + d) \proj{7/2,2}{7/2,2}+  \nonumber \\
&&+ \Delta \left( \proj{7/2,1}{7/2,1} + \proj{7/2,3}{7/2,3} \right) + \nonumber\\ 
& &+  \Omega e^{- i \omega t} \left[ \,
 S_{1,2} \left( \proj{7/2,2}{9/2,1} + \proj{7/2,1}{9/2,2}  \right) \right. + \nonumber \\
&   &  \phantom{+\Omega e^{- i \omega t}} +
S_{1,2}^* \left( \proj{7/2,2}{9/2,3} + \proj{7/2,3}{9/2,2} \right) + \nonumber \\
& & \phantom{+\Omega e^{- i \omega t}} + 
S_{1,1} \, \left( \proj{7/2,1}{9/2,1} + \proj{7/2,3}{9/2,3} \right) + \nonumber \\
&   & \phantom{+\Omega e^{- i \omega t}} +
 \left.  S_{2,2} \,  \proj{7/2,2}{9/2,2}  \ \right] \;  +\;  {\rm H.c.}
\label{eq:6levelsmodel}
\end{eqnarray}
Once we apply the unitary transformation 
$$\Gamma(t) = \exp [i \, d \left( |9/2,2\rangle \langle 9/2,2 | + |7/2,2\rangle \langle 7/2,2 | \right) t],$$ 
the three levels $\ket{9/2,k}$ become degenerate.
In case the three inequalities $|S_{i,j} \Omega| / (\delta - d) \ll 1$ are fulfilled,
it is possible to use second-order perturbation theory 
in order to develop an effective Hamiltonian describing 
the dynamics within the sub-manifold we are interested in, namely
\begin{eqnarray}
H_{pert} / \Omega^2 & = & - \left( \frac{|S_{1,1}|^2}{\delta-d} + \frac{|S_{1,2}|^2}{\delta} \right) \cdot \nonumber \\
&&\cdot [\proj{9/2,1}{9/2,1} + \proj{9/2,3}{9/2,3}] \nonumber \\
&  & - \left( \frac{|S_{2,2}|^2}{\delta - d} + 2 \frac{|S_{1,2}|^2}{\delta-2d}\right) \proj{9/2,2}{9/2,2}  \nonumber \\
&  & - \frac{S_{1,2}^{\hspace{0.05cm}2}}{\delta} \proj{9/2,3}{9/2,1} \;+\; H.c. \nonumber \\
&  & - \left[ \frac{ S_{1,2}^* \, S_{1,1}}{2} \left( \frac{1}{\delta - d} + \frac{1}{ \delta-2d} \right) e ^{i d t} \, + \right. \nonumber \\
&  & + \left. \frac{ S_{2,2}^* \, S_{1,2}}{2} \left( \frac{1}{\delta - d} + \frac{1}{ \delta} \right) e ^{i d t}  \right] \cdot \nonumber \\
& & \cdot \, [\proj{9/2,2}{9/2,1} + \proj{9/2,2}{9/2,3} ]   \;+\; H.c.
\label{eq:pert}
\end{eqnarray}
Remarkably enough, this Hamiltonian leads to the desired transfer rate of population 
from level $\ket{9/2,1}$ to $\ket{9/2,3}$, and viceversa. 
The main contribution is the direct coupling 
\begin{equation}
-J^{(1)}_{13} e^{i 2 \phi} = - \frac{|S_{1,2}|^2 \Omega^2 }{ \delta} e^{i 2 \phi}; \qquad
\phi = \arg S_{1,2}.
\label{eq:hoprate1}
\end{equation}
A second contribution, which in our system will prove to be not-negligible, 
comes from a sort of ``adiabatic elimination'' of the level $\ket{9/2,2}$, namely
\begin{equation}
-J^{(2)}_{13} = - \frac{ \bra{9/2,3}H_{pert}\ket{9/2,2} \, \bra{9/2,2} H_{pert} \ket{9/2,1}}
{\bra{9/2,2} H_{pert} \ket{9/2,2} - \bra{9/2,1} H_{pert} \ket{9/2,1} + d} .
\label{eq:hoprate2}
\end{equation}
Accordingly, we have derived the desired effective Hamiltonian where the Raman lasers assist the hopping of the physically meaningful $F=9/2$ levels, after the auxiliary $F=7/2$ bus states have been adiabatically eliminated. In the following sections, we shall address the range of validity of the approximations leading to this Hamiltonian, and compare it with the exact numerical investigation of the initial Hamiltonian~(\ref{eq:6levelsmodel}).

We want to stress here that even if  the integrals in the definition~(\ref{eq:S}) of the $S_{kk'}$ can be complex numbers, this does not have any physical influence on this proposal. Indeed, even if the effective coupling between neighboring main sites $-J$ was complex, its spatially uniform phase  can be gauged away with a space-dependent unitary  transformation (even in the case of periodic boundary conditions). Conversely, the non-uniform phase coming from the $e^{i \mathbf{q}_t \cdot \mathbf x_k}$ factor, which arises when $\mathbf q_t$ is not parallel to the direction of the hopping it assists, cannot be gauged away even in presence of open boundary conditions. 
Such a phase, which is not related to the fact that the integrals in~(\ref{eq:S}) are complex, can be used to simulate an external uniform magnetic field~\cite{JakschZoller, GerbierDalibard}.
Finally, we underline that in our setup, where the tunneling along each axis is induced by lasers propagating parallel to the axis itself, both complex phases can be gauged away. In order to simulate a magnetic field, therefore, one should move slightly away from this configuration and engineer a Raman coupling whose effective transmitted momentum does not run parallel to the links of the lattice. We will not consider this situation in this article because the models of interest in section~\ref{sec:proposals} do not require such space-dependent phase.

\subsection{Range of validity of the ``6-Level Model''}

The presented ``6-level model'' strongly relies on two approximations:
\begin{enumerate}
\item \label{issue2} considering the bands of the lattice as being flat;
\item \label{issue1} neglecting delocalized higher-energy free states.
\end{enumerate}
If these approximations are not justified for a given experimental configuration, spurious population transfers  to next-neighboring sites would arise. 

The approximation~(\ref{issue2}) is required to fulfill the core idea of the proposal, namely the adiabatic elimination of the intermediate level. This is demonstrated with a model which considers only a subset of the Hilbert space spanned by the real eigenstates of the Hamiltonian (Bloch functions), considering just three of their linear combinations (the Wannier functions $w_{k=1} (\mathbf x)$, $w_2(\mathbf x)$ and $w_3(\mathbf x)$).
This is equivalent to approximating the dispersion laws of the band as being flat, neglecting  thus possible curvature effects, and is  legitimated as long as the width of the band is much smaller than the detuning of the transition $\delta - d$. In case the degeneracy of the Bloch functions cannot be assumed, all the Bloch functions should be considered in order to quantitatively estimate the spurious effects cited above.
In general, this issue sets a trade-off for the relative depth $\xi$ of the secondary lattice in (\ref{eq:potential}): on one hand, a shallow lattice ($\xi < 1$) is desirable because the Wannier function of the intermediate minimum $w_{k=2}(\mathbf x)$ is not strongly localised and laser-induced transitions are favored ($|S_{1,2}| \sim |S_{1,1}|$). On the other hand, the more the wavefunction is delocalized, the more the band bends, eventually becoming  parabolic at $\mathbf k =0$  with a bandwidth comparable to the detuning. In our numerical simulations we consider $\xi = 1$, which is a reasonable middle-way.

Regarding the issue~(\ref{issue1}),  higher-energy bands could become  important in the presence of intense Raman transitions $\Omega$ and large detunings $\delta-d$,  which couple them to the lowest-band states. The presented analytical and numerical studies do not take into account these effects since they consider only three Wannier functions and effectively only two bands, 
even though including bands with localized Wannier functions would just imply a renormalization of the numerical coefficients $S_{k,k'}$. A different problem is the case of high-energy strongly parabolic bands, whose Wannier functions are not strongly localized. The effect of such states is not considered by our model, which is that of spreading population among many next- and further-neighboring main sites.
From an experimental point of view, we expect a trade-off to arise
between a large detuning regime, allowing powerful lasers and strong effective couplings with noisy spurious population transfers, and a small detuning one, with clean but small couplings. 

\begin{table}[p]

\caption{Parameters used for the numerical simulation of the diagonal hopping in subsection~\ref{subsec:diaghop}. We list the numerical values of all the main parameters characterizing the atomic transitions and the Raman couplings. The first Raman coupling induces the hopping of the $F=9/2$, $m_F =9/2$ whereas the second one addresses the $F=9/2$, $m_F = 7/2$ (such states were however not considered in the simulation).}
\label{table:diaghoprun}

\begin{center}
\begin{tabular}{|cc|cc|}
\br
Level: $\ket{F, m_F, k}$ & Energy & Parameters &\\
\mr
$\ket{ 9/2; 9/2;  1}$ & $ g_F \mu_B B m_F $ & $\hfsplit $ & $1.285$ GHz \\
$\ket{ 9/2; 9/2; 2}$ & $ g_F \mu_B B m_F +d$ & $\mu_F B$ & $40$ MHz\\
$\ket{ 9/2; 9/2; 3}$ & $ g_F \mu_B B m_F $ & $d$  & $69.160$ kHz\\
$\ket{ 7/2; 7/2; 1}$ & $\hfsplit - g_F \mu_B B m_F $ & $S_{1,1}$ & $0.46$ \\
$\ket{ 7/2; 7/2; 2}$ & $\hfsplit - g_F \mu_B B m_F +d$ & $S_{1,2}$ & $0.07 + i 0.13$ \\
$\ket{ 7/2; 7/2; 3}$ & $\hfsplit - g_F \mu_B B m_F $ & $S_{2,2}$ & $0.16$ \\
\br
\end{tabular}

\begin{tabular}{|c|c|c|}
\br
\# Raman & $\Omega$ & $\omega$ \\
\mr
1 & 49.5 kHz & $E_{|7/2;7/2;2\rangle} - E_{|9/2;9/2;1\rangle} - 300$ kHz \\
2 & 49.5 kHz & $E_{|7/2;5/2;2\rangle} - E_{|9/2;7/2;2\rangle}  - 300$ kHz \\
\br
\end{tabular}

\begin{tabular}{|cc|cc|}
\br
$J_{13}^{(1)}$ & $J_{13}^{(2)}$ & Estimated T & Numerical T \\
\mr
$176$ Hz & $17$ Hz & $0.018$ s & $0.017$ s \\
\br
\end{tabular}
\end{center}

\caption{Parameters used for the numerical simulation of the non-diagonal hopping in subsection~\ref{subsec:nondiag}. We list the numerical values of  the main parameters characterizing atomic transitions and  Raman couplings. The reported frequencies of the Raman couplings are approximate because some additional tuning is needed to compensate the different Stark shift for states with $k=1$ and $k=3$ arising due to Raman dressing in presence of staggering. Perfect matching the atomic transition becomes difficult and imperfections are responsible for the not clean population transfer in figure~\ref{fig:timeevol_NDH}. Larger staggering values would help.}
\label{table:NONdiaghoprun}

\begin{center}
\begin{tabular}{|cc|cc|}
\br
Level: $\ket{F, m_F, k}$ & Energy & Parameters &\\
\mr
$\ket{ 9/2; 9/2;  1}$,$\quad$ $\ket{ 9/2; 7/2;  1}$& $ g_F \mu_B B m_F $ & $\hfsplit $ & $1.285$ GHz \\
$\ket{ 9/2; 9/2; 2}$,$\quad$ $\ket{ 9/2; 7/2;  2}$ & $ g_F \mu_B B m_F +d $ & $\mu_F B$ & $40$ MHz\\
$\ket{ 9/2; 9/2; 3}$,$\quad$ $\ket{ 9/2; 7/2;  3}$ & $ g_F \mu_B B m_F +15$ kHz & $d$ & $69.160$ kHz\\
$\ket{ 7/2; 7/2; 1}$,$\quad$ $\ket{ 7/2; 5/2;  1}$ & $\hfsplit - g_F \mu_B B m_F $ & $S_{1,1}$ & $0.46$ \\
$\ket{ 7/2; 7/2; 2}$,$\quad$ $\ket{ 7/2; 5/2;  1}$ & $\hfsplit - g_F \mu_B B m_F +d$ & $S_{1,2}$ & $0.07 + i 0.13$ \\
$\ket{ 7/2; 7/2; 3}$,$\quad$ $\ket{ 7/2; 5/2;  1}$ & $\hfsplit - g_F \mu_B B m_F +15$ kHz & $S_{2,2}$ & $0.16$ \\
\br
\end{tabular}

\begin{tabular}{|c|c|c|}
\br
\# Raman & $\Omega$ & $\omega$ \\
\mr
1 & 49.5 kHz & $\sim E_{|7/2; 7/2; 2 \rangle} - E_{|9/2; 9/2; 1 \rangle} - 300$ kHz \\
2 & 49.5 kHz & $\sim E_{|7/2; 5/2; 2\rangle} - E_{|9/2; 7/2; 1 \rangle} - 300$ kHz \\
3 & 49.5 kHz & $\sim E_{|7/2; 7/2; 2 \rangle} - E_{|9/2; 7/2; 3 \rangle} - 300$ kHz \\
\br
\end{tabular}

\end{center}

\end{table}

\subsection{Diagonal Hopping Operator} \label{subsec:diaghop}

\begin{figure}[t]
\begin{center}

\includegraphics[width=\columnwidth]{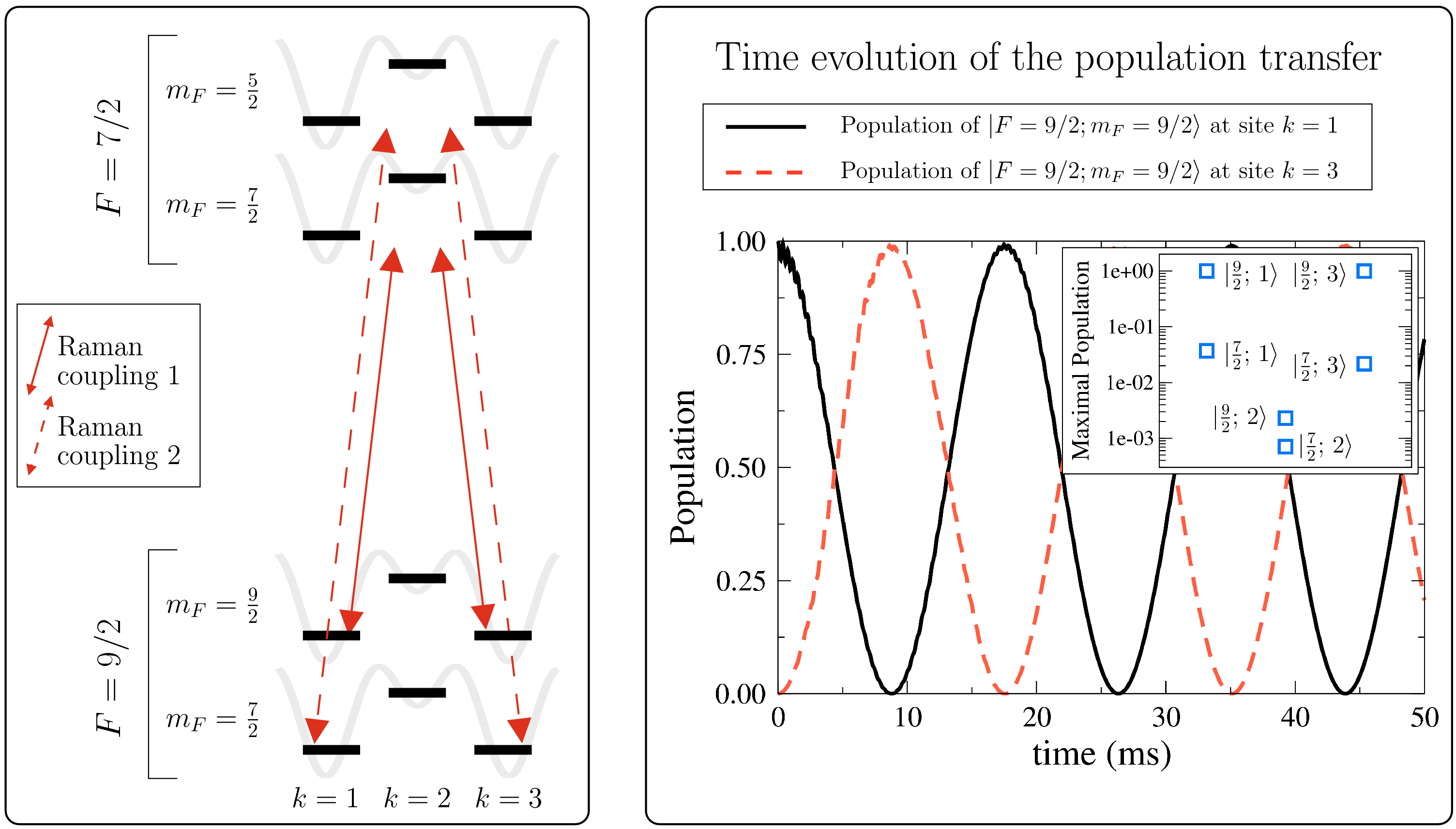}
\caption{
Left: sketch of the scheme proposed for the realization of a diagonal hopping operator (Energies are not in scale). Raman coupling $1$ ($2$) connects the $\ket{F=9/2, m_F=9/2}$ ($\ket{F=9/2, m_F=7/2}$) states to their auxiliary state. Detuning allows independent control of the hopping rates.
Right: exact time-evolution of the ``6-levels model''~(\ref{eq:6levelsmodel}), showing the coherent population transfer between sites $1$ and $3$ of the spin state $\ket{F=9/2, m_F=9/2}$.
The parameters used are listed in table~\ref{table:diaghoprun}. The inset shows the maximal populations of the six considered levels labelled $\ket{F,k}$ as in~(\ref{eq:6levelsmodel}) and shows that only a small fraction of the population is lost in auxiliary levels. }
\label{fig:hoppingscheme}
\label{fig:timeevol_DH}
\end{center}
\end{figure}

We now explicitly study the possibility of realising a diagonal tunneling operator.
We numerically simulate the Hamiltonian~(\ref{eq:6levelsmodel}) with a simple Runge-Kutta algorithm. 
We did not include in the simulation hyperfine states different from $\ket{F=9/2, m_F = 9/2}$ and $\ket{F=7/2, m_F=7/2}$ because they are strongly detuned from those we are considering. However, for completeness, we include the presence of a second Raman coupling which would be needed to induce the hopping of $\ket{F=9/2; \, m_F=7/2}$ and check that it is unimportant.

We show in figure~\ref{fig:timeevol_DH}  the numerical results. The realistic parameters used in this simulation are listed in table~\ref{table:diaghoprun}. The population is coherently transferred between two neighboring levels and only a negligible fraction is lost in auxiliary states.
Regarding the validity of the ``6-levels model'', for the lattice considered here, the bandwidths of the two bands are respectively $0.2$ kHz and $9.1$ kHz, which should be compared with the considered detuning of $300$ kHz. In these and the following simulations, the employed numerical values have only an illustrative purpose and other regimes could be considered.

Therefore, these results confirm  the plausibility of our scheme to induce a laser-assisted tunneling between the atoms sitting in the main minima of the optical lattice. To make the simulation toolbox reacher, we now address the possibility to control  a spin-dependent hopping process.

\subsection{Non-Diagonal Hopping Operator} \label{subsec:nondiag}

\begin{figure}[t]
\begin{center}
\includegraphics[width=\columnwidth]{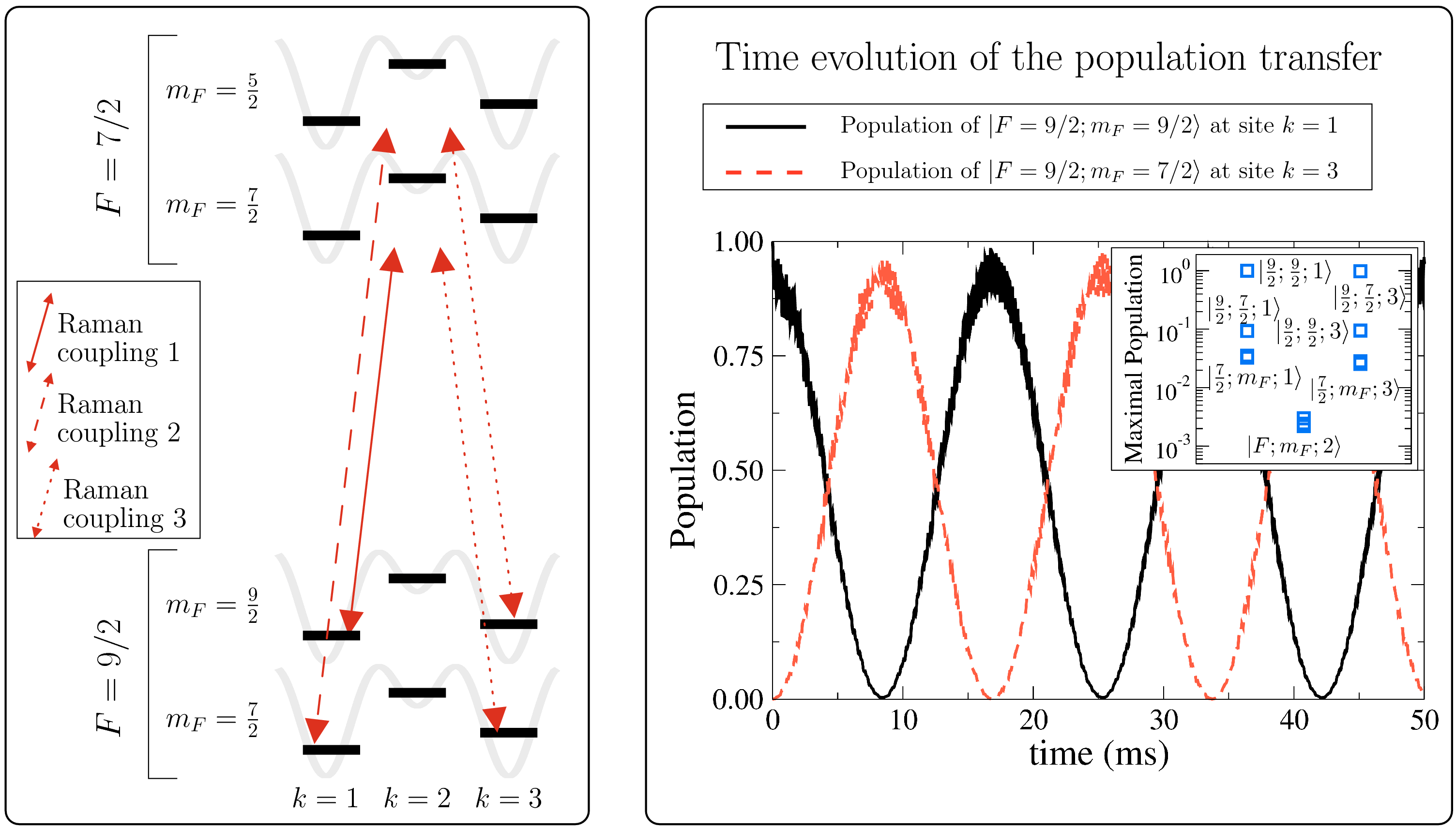}
\caption{
Left: sketch of the scheme proposed for the realization of a non-diagonal hopping operator (Energies are not in scale). Raman couplings $1$, $2$ and $3$ connect the $\ket{F=9/2, m_F=9/2}$  and $\ket{F=9/2, m_F=7/2}$ states to the auxiliary states. Detuning allows independent control of the hopping rates.
Right: exact time-evolution of the ``12-levels model'' introduced in subsection~\ref{subsec:nondiag} and generalizing~(\ref{eq:6levelsmodel}). We show the coherent population transfer between sites $1$ and $3$ of the spin state $\ket{F=9/2, m_F=9/2}$ and $\ket{F=9/2, m_F=7/2}$.
The parameters used are listed in table~\ref{table:diaghoprun}. The inset shows the maximal populations of the twelve considered levels labelled $\ket{F,m_F,k}$ and shows that only a fraction of the population is lost in auxiliary levels.
}
\label{fig:timeevol_NDH}
\end{center}
\end{figure} 

In order to study the realisation of the non-diagonal hopping operator, we consider an enlarged 12-levels model, which is a generalization of the previous one taking into account more hyperfine states. We want now to transfer population between the manifolds $\ket{F=9/2, m_F = 9/2}$ and $\ket{F=9/2, m_F = 7/2}$ and consider as auxiliary states $\ket{F=7/2, m_F = 7/2}$ and $\ket{F=7/2, m_F = 5/2}$.

A big issue which must be solved to engineer such a hopping is the arousal of undesired spin-flipping terms induced by the laser. In this paper we consider the possibility of staggering the lattice with an additional optical field, in order to lift the degeneracy between the different sites of the optical lattice, in the same fashion of~\cite{GerbierDalibard}. Such a staggering can be done also in three dimensions since the cubic lattice is bipartite, and we consider staggering values of $10-15$ kHz.

Figure~\ref{fig:timeevol_NDH} sketches the experimental scheme we have in mind and
shows the exact time evolution of the population transfer between the two levels of $F=9/2$ in two neighboring sites. Interestingly enough, we show a flip of the Zeeman spin during the tunneling process, and thus obtain the promised spin-dependent hopping operator. The parameters of the simulation can be found in table~\ref{table:NONdiaghoprun}.

\section{From a Spin-Dependent Hopping Operator to a Quantum Simulator\label{sec:critical}}

In the previous section, we discussed how the superlattice geometry could be used to create non-trivial hopping operators on each link. Here we want to assemble these ingredients and discuss how to use them to engineer interesting quantum simulator in arbitrary dimensions, considering the advantages and disadvantages of the proposal.

First of all, we stress that the lasers needed to engineer the hopping along one direction must transfer momentum along that same direction. Therefore, just by controlling the beam propagation directions, we can tailor different tunneling operators along each axis. This is an important feature which will be largely exploited in the proposals listed in section~\ref{sec:proposals}.
This kind of \emph{directionality selection rule} is also responsible for avoiding the population of higher-order minima which do not lie on the edges of the unit lattice cell. Since they are not connected to the main minima by a line parallel to a cartesian axis, we do not consider any momentum transfer along such direction, and thus the formal orthogonality of the Wannier functions localized in those minima is never  lifted.

Due to the very general formulation of the superlattice potential, the setup is well-suited also to work in two and one dimensions. Moreover in 1D it is possible to align the magnetic field splitting the hyperfine sublevels with the optical lattice. In this case, it is possible to use specific light polarization to selectively couple different atomic levels. 
As a short-term goal, it would be very interesting to understand what is the most interesting physics which could be simulated in a one-dimensional system, where the presence of more symmetries could lower the experimental intricacies. We partially address this question in section~\ref{sec:proposals}, where we argue that several one-dimensional topological phases could be realized.
In a three- (two-)dimensional case, with the magnetic field aligned with the diagonal of the cube (square), the polarization of the light cannot be exploited, only energy-based selection rules are reliable.

Unfortunately, energy-based selection rules are not enough to prevent, in the case of spin-flipping operators, spurious on-site spin-flipping couplings. We proposed to solve this issue by staggering the optical lattice, i.e. lifting the degeneracy of the lattice sites of some tens of kHz (see also the discussion in reference~\cite{GerbierDalibard}). One should also mention that  the diagonal hopping operators can still be engineered in presence of such staggering, with the only additional issue of using two Raman couplings (as in the non-diagonal case) to match the energy difference between sites.

Each of the matrix elements of the hopping operators is realized via an effective four-photon process. This means that the spin of the atom can be flipped 
of at most $|\Delta m_F| = 4$: a careful analysis is needed in case one is interested in simulating a theory with more than 4 fields, because some hopping matrix elements might become avoided.

Finally, the description given in this proposal is essentially at the single-particle level where no many-body effects have been used. As a consequence, the proposal works both for bosons and fermions, which is a valuable result. 
The proposals of section~\ref{sec:proposals} assume the possibility of switching off atomic interactions with Feshbach resonances. Nonetheless, it would be of the utmost  interest to study how interactions could fit into this picture, and to explore the rich variety of models that could emerge. We leave this problem for future work.

We now conclude the part of the article devoted to the description of the experimental setup. We believe to have provided the relevant results supporting the initial claim that it was indeed possible to realize a system whose low-energy structure is described by Hamiltonian~(\ref{eq:claim}).

\section{Applications of the Quantum Simulator\label{sec:proposals}}

In this section, we would like to discuss the use of the described setup as a quantum simulator.
We will address a range of lattice field theories for relativistic fermions~\cite{kogut_rmp}, 
and explore exotic phases of matter known as topological insulators~\cite{top_insulators_rev}. In general,
the task of a purpose-based quantum simulator is to realize a system described by an effective Hamiltonian $H_{\mathrm{eff}}$ that reproduces faithfully the properties of the model to be simulated. In our case, this model corresponds to relativistic lattice fermions $H_{\mathrm{rel}}$, or  topological insulators $H_{\mathrm{top}}$. The resource to accomplish such a task is the microscopic control over the superlattice setup, which we have argued previously to be described by the Hamiltonian~(\ref{eq:claim}), rewritten here  for reading convenience,
\begin{equation*}
H_{\mathrm{sys}}=
\sum_{\mathbf{r}\bm{\nu}}\sum_{\tau\tau'}
t_{\nu}c^{\dagger}_{\mathbf{r}+\bm{\nu}\tau'}[U_{\bm{\nu}}]_{\tau'\tau}c_{\mathbf{r}\tau}+\sum_{\mathbf{r}}\Omega c^{\dagger}_{\mathbf{r}\tau'}[\Lambda]_{\tau'\tau}c_{\mathbf{r}\tau}+\mathrm{H.c.} 
\end{equation*}
The main objective in the following subsections is to control and manipulate:
\begin{enumerate}
\item the optical lattice dimension $D$;
\item the tunneling strengths $t_{\nu}$;
\item the spin-dependent hopping operators $U_{\bm{\nu}}$;
\item the on-site Raman transitions $\Omega,\Lambda$;
\end{enumerate}
such that the Hamiltonian of equation~(\ref{eq:claim}) simulates the desired physics, namely $$H_{\mathrm{sys}}(\{t_{\nu},U_{\mathbf{r}\bm{\nu}},\Lambda,\Omega\}) \;\to\; H_{\mathrm{eff}}\approx H_{\mathrm{rel}},H_{\mathrm{top}}.$$

From a condensed-matter perspective, exotic phases are frequently associated to strongly-correlated regimes and many-body interactions. There are however distinguished exceptions to this paradigm, such as graphene~\cite{graphene_rev} and topological insulators~\cite{top_insulators_rev}, where quadratic fermionic Hamiltonians contain a wealth of non-trivial phenomena. In the case of graphene, a two-dimensional layer of graphite, the low-energy carriers can be described by emerging relativistic fermions without mass. On the other hand, topological insulators are exotic holographic phases with an insulating bulk, and a peculiar  boundary that hosts robust conducting modes protected by topology arguments. In both cases, the transport properties of the material differ significantly from the standard solid-state theory. In the subsections below, we show how the Hamiltonian~(\ref{eq:claim}) serves as a versatile simulator of these two interesting phases of matter. 

\subsection{The Zoo of Relativistic Lattice Fermions}

The properties of a relativistic spin-$1/2$ fermion with mass $m$  are described by the famous Dirac Hamiltonian~\cite{Peskin}
\begin{equation}
\label{dirac_ham}
H=\int \mathrm{d} \mathbf r \; \Psi(\mathbf{r})^{\dagger}H_{\mathrm{DI}}\Psi(\mathbf{r}),\qquad H_{\mathrm{DI}}=c\bm{\alpha}\cdot\mathbf{p}+mc^2\beta;
\end{equation}
where $\alpha_{\nu},\beta$ are the so-called Dirac matrices fulfilling a Clifford algebra, $\{\alpha_{\nu},\alpha_{\mu}\}=2\delta^{\nu\mu}$, $\{\alpha_{\nu},\beta\}=0$, and $c$ stands for the speed of light. Here, $\Psi(\mathbf{r})$ is the $N_{D}$-component fermionic field operator, where $N_D=2$ for one and two spatial dimensions, and $N_D=4$ for three spatial dimensions. Our objective now is to construct an effective Hamiltonian starting from equation~(\ref{eq:claim}) that closely resembles the relativistic field theory in equation~(\ref{dirac_ham}). The underlying setup consists of a gas of ultracold $^{40}$K atoms, which is a  non-relativistic system; nonetheless we can design it as a quantum simulator of relativistic particles by exploiting the quantum statistics and a peculiar engineerable Fermi surface.

\subsubsection*{Naive Massless or Massive Dirac Fermions.} 

The idea is to engineer translationally invariant hopping operators, $U_{\nu}=\ee^{\ii \phi_{\nu}A_{\nu}}$,  according to the SU($N_D$) group, where $N_D$ has been defined above. For the particular choices specified in table~\ref{naive_table}, one finds that the Hamiltonian in equation~(\ref{eq:claim}) in momentum space becomes
\begin{equation}
\label{mom_ham}
H=\sum_{\mathbf{k}\in\mathrm{BZ}}\Psi^{\dagger}_{\mathbf{k}}\left(\sum_{\nu}2t_{\nu}\cos\phi_{\nu}\cos k_{\nu}\mathbb{I}+2t_{\nu}\sin\phi_{\nu}\sin k_{\nu}\alpha_{\nu}\right)\Psi_{\mathbf{k}},
\end{equation}
where $\Psi_{\mathbf{k}}$ is a multicomponent fermi operator that contains the different $N_D$ hyperfine levels involved in the simulation, and ${\bf k}$ is defined within the first Brillouin zone BZ. One readily observes that 
there are certain regimes, the so-called $\pi$-flux phases $\phi_{\nu}=\pi/2$, where the energy spectrum develops $\mathcal N_D=2^D$ degeneracy points $\mathbf{K}_{\mathbf{d}}$ where the energy bands touch $\epsilon(\mathbf{K}_{\mathbf{d}})=0$. Around these points $\mathbf{K}_{\mathbf{d}}=(d_{x}\pi,d_y\pi,d_z\pi)$, where $d_{\nu}\in\{0,1\}$ is a binary variable, the low-energy excitations of the $^{40}$K Fermi gas are described by the effective Hamiltonian
\begin{equation}
\label{naive_df}
H_{\mathrm{eff}}=
\sum_{\mathbf{d}}
\sum_{ \mathbf{p}_{\mathbf{d}}}
\Psi^{\dagger}( \mathbf{p}_{\mathbf{d}}) \, H_{\mathrm{DI}}^{\mathbf{d}} \, \Psi( \mathbf{p}_{\mathbf{d}}),
\qquad  
H_{\mathrm{DI}}^{\mathbf{d}}(\mathbf{p}_{\mathbf{d}})=c \bm{\alpha}^{\mathbf{d}} \cdot \mathbf{p}_{\mathbf{d}},
 \end{equation}
where $\mathbf{p}_{\mathbf{d}}=\mathbf{k}-\mathbf{K}_{\mathbf{d}}$ represents the momentum around the degeneracy points,  $(\bm{\alpha}^{\mathbf{d}})_{\nu}=(-1)^{d_{\nu}}\alpha_{\nu}$ are the Dirac matrices listed in table~\ref{naive_table}, and $c=2t_{x}=2t_{y}=2t_{z}$ is the Fermi velocity that plays the 
role of an effective speed of light. Therefore, the Fermi surface of the half-filled gas consists of a set of isolated points, the so-called Dirac points, and the low-energy excitations around those points behave according to the Hamiltonian of massless Dirac fermions in equation~(\ref{naive_df}). 
As occurs with graphene~\cite{graphene_rev}, or other cold-atom systems~\cite{Hou, Lim, GoldAnom, Lepori, Celi, dirac_fermions}, one finds a table-top experiment of relativistic quantum field theories. 

Let us note that we obtain an even number of relativistic-fermion species, each located around a different Dirac point (i.e. $\mathcal N_1=2$ for one dimension, $\mathcal N_2=4$ for two dimensions, and $\mathcal N_3=8$ for three dimensions). This doubling of fermionic species is a well-known phenomena in lattice gauge theories~\cite{kogut_rmp}, where the fermions in equation~(\ref{naive_df}) would correspond to the so-called {\it naive Dirac fermions}~\cite{naive_fermions}. As predicted by the Nielsen-Ninomiya theorem~\cite{fermion_doubling}, this doubling cannot be avoided without breaking an underlying symmetry:
\begin{itemize}
\item for $D$ odd $\, \{H_{\mathrm{DI}}^{\mathbf{d}},\Gamma_{\mathrm{1}}\}=0\,$ is an involution known as  chiral symmetry;
\item for $D$ even $\Gamma_{\mathrm{2}}^{\dagger} \,\left[ H_{\mathrm{DI}}^{\mathbf{d}}(-\mathbf{p}_{\mathbf{d}})\right]^*  \,\Gamma_{\mathrm{2}}=H_{\mathrm{DI}}^{\mathbf{d}}(\mathbf{p}_{\mathbf{d}})$ is an antiunitary symmetry known as time-reversal (see table~\ref{naive_table}). 
\end{itemize}

\begin{table}
\caption{ Quantum simulator of naive Dirac fermions. Each translationally-invariant hopping operator $U_{\bm{\nu}}$ is expressed in terms of Pauli matrices $\{ \sigma_{x}, \sigma_y, \sigma_z\}$, and a set of dimensionless fluxes $\{ \phi_{1}, \phi_2, \phi_3 \}$. We also list a particular representation of the Dirac matrices $\alpha_{\nu},\beta$ for different spatial dimensions $d$, together with the important symmetries of the Hamiltonian that depend on $\Gamma$.}\label{naive_table}

\begin{center}

\begin{tabular}{|c|c|c|c|}
\br
$ D$ & $U_{\mathbf a_1}$ & $U_{\mathbf a_2}$ &$U_{\mathbf a_3}$  \\
\mr
1 & $\ee^{\ii\phi_1 \sigma_x}$& &  \\
2& $\ee^{\ii\phi_1 \sigma_x}$&  $\ee^{\ii\phi_2\sigma_y}$ & \\
3 & $\ee^{\ii\phi_1 \sigma_z\otimes\sigma_x}$&$\ee^{\ii\phi_2\sigma_z\otimes\sigma_y}$ &$\ee^{\ii\phi_3\sigma_z\otimes\sigma_z}$ \\
\br
\end{tabular}

\vspace{0.25cm}

\begin{tabular}{|c|c|c|c|c|c|}
\br
$ D$ & $\alpha_x$ & $\alpha_y$ & $\alpha_z$ & $\beta$ & $\Gamma$\\
\mr
1 & $\sigma_x$&&&$\sigma_z$ & $\ii\beta\alpha_x$\\
2 & $\sigma_x$&$\sigma_y$& &$\sigma_z$ & $\ii\sigma_y$\\
3 & $\sigma_z\otimes\sigma_x$ & $\sigma_z\otimes\sigma_y$ &  $\sigma_z\otimes\sigma_z$&$\sigma_x\otimes\mathbb{I}_2$ & $-\ii\alpha_x\alpha_y\alpha_z$ \\
\br
\end{tabular}

\end{center}

\end{table}

According to these results, we have a versatile quantum simulator of {\it massless Dirac fermions} in any spatial dimension $D=1,2,3$. In $D=1,2$ they coincide with the famous {\it Weyl fermions}, whereas in $D=3$ they contain a couple of Weyl fermions with opposed helicities. Note that this scheme can also be extended so as to simulate exotic Weyl fermions of any arbitrary spin $s$~\cite{weyl}. Besides, our quantum simulator also allows us to make these fermions massive, thus reaching the desired Hamiltonian in equation~(\ref{dirac_ham}). The idea is to control the on-site Raman transitions such that $\Lambda=\beta$ listed in table~\ref{naive_table}. In such case, the Rabi frequency plays the role of the mass $mc^2=2\Omega$, and the effective Hamiltonian in equation~(\ref{naive_df}) becomes 
\begin{equation}
H_{\mathrm{DI}}^{\mathbf{d}}(\mathbf{p}_{\mathbf{d}})=c\bm{\alpha}^{\mathbf{d}}\cdot\mathbf{p}_{\mathbf{d}}+mc^2\beta.
\end{equation}
Therefore, this quantum simulator can explore both the non-relativistic and the ultra-relativistic limits of the theory.

\subsubsection*{Wilson and Kaplan Fermions.} 

From a lattice gauge theory perspective, the additional fermions  around  $\mathbf{K}_{\mathbf{d}}\neq \mathbf{0}$ are spurious doublers 
that modify the physics at long wavelengths. A partial solution is to give the doublers a very large mass $m_{\mathbf{K}_{\mathbf{d}}}c^2$, so that they effectively decouple from the low-energy physics of the Dirac fermion at $\mathbf{K}_{\mathbf{d}}= \mathbf{0}$, namely $m_{\mathbf{K}_{\mathbf{d}}} \gg m_{\mathbf{K}_{\mathbf{0}}}$. We must find a way to engineer a momentum-dependent mass that differs from the global on-site Raman mass discussed above. By combining the laser-assisted tunneling listed in table~\ref{naive_table}, with the additional terms $\tilde{U}_{\bm{\nu}}=\ii\ee^{\ii \varphi_{\nu}\beta}$~\cite{MazzaRizziPRL}, the momentum-space  Hamiltonian $H$ in equation~(\ref{mom_ham}) becomes $H+\tilde{H}$, where
\begin{equation}
\tilde{H}=\sum_{\mathbf{k}\in\mathrm{BZ}}\Psi^{\dagger}_{\mathbf{k}}\left(\sum_{\nu}2\tilde{t}_{\nu}\cos\varphi_{\nu}\sin k_{\nu}\mathbb{I}-2\tilde{t}_{\nu}\sin\varphi_{\nu}\cos k_{\nu}\beta\right)\Psi_{\mathbf{k}},
\end{equation}
where $\tilde{t}_{\nu}$ are the additional laser-assisted tunneling strengths. Once more, for the $\pi$-flux phases $\varphi_{\nu}=\pi/2$, the effective Hamiltonian in equation~(\ref{dirac_ham}) is modified into
\begin{equation}
\label{wilson_mass}
 H_{\mathrm{DI}}^{\mathbf{d}}(\mathbf{p}_{\mathbf{d}})=c\bm{\alpha}^{\mathbf{d}}\cdot\mathbf{p}_{\mathbf{d}}+m_{\mathbf{K}_{\mathbf{d}}}c^2\beta,\hspace{1ex} m_{\mathbf{K}_{\mathbf{d}}}=m-\sum_{\nu}(-1)^{d_{\nu}}m_{\nu},
\end{equation} 
where   $m_{\nu}c^2=2\tilde{t}_{\nu}$. Let us emphasize that our quantum simulator provides a complete control over the different masses, since $m$ depends on the on-site Raman transition strengths, whereas $m_{\nu}$ 
depends on the assisted-hopping strength, and thus on the laser power. In particular, when these parameters fulfill $\sum_{\nu}m_{\nu}=m$ (i.e. $m_x=m$ for $D=1$, $m_x+m_y=m$ for $D=2$, and $m_x+m_y+m_z=m$ for $D=3$), we obtain a single massless Dirac fermion at $\mathbf{K}_{\mathbf{d}}= \mathbf{0}$, whereas the remaining doublers have been boosted to much higher energies. We thus achieve a quantum simulator of the so-called {\it Wilson fermions} of any spatial dimension $D=1,2,3$~\cite{wilson}.

We note that this decoupling between a single massless Dirac fermion and its doublers is not in conflict with the Nielsen-Ninomiya theorem since the introduced mass terms explicitly break the aforementioned symmetries. This is particularly important in odd dimensions, where the theory does not preserve chiral symmetry,   a fundamental concept in  the standard model  classifying right/left-handed particles $\Gamma_1\Psi=\pm\Psi$. To preserve such symmetries, the concept of {\it Kaplan fermions} arises~\cite{kaplan}, namely massless Dirac fermions bound to a lower-dimensional domain wall located at $\mathbf{r}_{\bot}^*$ where the Wilson mass gets inverted $m_{\mathbf{K}_{\mathbf{d}}}=-|m|+2|m|\theta(\mathbf{r}_{\bot}-\mathbf{r}_{\bot}^*)$.  Since we have a complete experimental access to the parameters of the Wilson mass, it is also possible to tune them such that $\sum_{\nu}m_{\nu}>m$, and thus the mass $m_{\mathbf{K}_{\mathbf{d}}}<0$ gets inverted, and one gets a lower-dimensional massless fermion bound to the region where this mass inversion takes place.

Let us close this subsection by underlying the versatility of our setup as a quantum simulator of a diverse set of relativistic lattice fermions. Not only can we implement massless Dirac fermions of any dimensionality, thus exploring their connection to Weyl fermions, but we can also control their mass. This leads us to the concept of massive Dirac fermions, and the notorious  Wilson and Kaplan fermions  dealing with the fermion doubling problem.  Interestingly enough, the physics behind these high-energy particles is intimately related to the materials known as topological insulators~\cite{top_insulators_rev}, which are the subject of the following subsection. In fact, it is always possible to find a Kaplan-fermion representative within each class of topological insulators~\cite{dim_reduction}.

\subsection{A Toolbox for Topological Insulators}

Topological insulators correspond to fermionic gapped phases of matter that are insulating in the bulk but allow a robust transport along the boundaries~\cite{top_insulators_rev}. This robustness   is due to the existence of gapless edge excitations which are protected against disorder by topological arguments (i.e. they avoid Anderson localization~\cite{anderson}, and thus transport charge even in the presence of strong disorder).  The paradigmatic example of a topological insulator is the integer quantum Hall effect (IQHE)~\cite{top_insulators_rev}, a two-dimensional electron gas subjected to a strong magnetic field that displays a robust quantization of the transverse conductivity $\sigma_{xy}=n \cdot e^2/h$, where $n\in\mathbb{Z}$ is related to the topological invariant known as the Chern number~\cite{thouless}. In this case, chiral electrons bound to the one-dimensional edges of the sample avoid back-scattering processes, and are thus immune to  disorder~\cite{qhe_edges}. Remarkably enough, the IQHE is only one instance of a large list of topological insulating phases.  Each class can be characterized by a set of discrete fundamental symmetries~\cite{top_ins_table}, and a certain topological invariant, see table~\ref{ti_table}. Note that we have excluded the topological superfluids from this table, since their quantum simulation would require a pairing mechanism, and thus goes beyond the scope of this work~\cite{top_superfluids, Duan}. We now discuss how our quantum simulator can reproduce the properties of many of these fascinating phases of matter following two possible strategies.

\begin{table}

\caption{Periodic table of topological insulators. The underlying quadratic  Hamiltonians $H=\sum_{\alpha\beta}\Psi^{\dagger}_{\alpha}\mathbb{H}_{\alpha\beta}\Psi_{\beta}+\mathrm{H.c.}$, where $\alpha,\beta$   represent the lattice sites and the internal states of the fermion, can be classified according to the fundamental symmetries of {\it time-reversal} $\mathcal{T}$, {\it charge conjugation} $\mathcal{C}$, and the combination of both $\mathcal{S}=\mathcal{T}\mathcal{C}$. The values $T=0$, $C=0$, $S=0$, are assigned to Hamiltonians that break the symmetry, whereas $T=\pm 1$, $C=\pm 1$, $S=1$ correspond to symmetry-preserving Hamiltonians, where $\mathcal{T}^2=\pm 1$, $\mathcal{C}^2=\pm 1$, $\mathcal{S}^2=+1$. There are six possible combinations for non-interacting fermionic Hamiltonians (and another four for pairing fermionic Hamiltonians), which lead to the classes listed in the first column. For each dimension $d$, there are three possible topological insulators among all these classes, and they are classified according to the {\it integer} $\mathbb{Z}$ or {\it binary} $\mathbb{Z}_2$ nature of a certain topological invariant. In the column labelled $QS$, we list the particular instances that can be simulated with our super-lattice based quantum simulator. }
\label{ti_table}
\begin{center} 

\begin{tabular}{|c|c|c|c|c|}
\br
\textbf{Class} & \textbf{Name} & $T$ & $C$ & $S$ \\
\mr
$\mathrm{A}$ & Unitary & $0$ & $0$ & $0$ \\
$\mathrm{AIII}$ & Chiral unitary & $0$ & $0$ & $1$ \\
$\mathrm{AI}$ & Orthogonal & $+1$ & $0$ & $0$ \\
$\mathrm{BDI}$ & Chiral Orthogonal & $+1$ & $+1$ & $1$ \\
$\mathrm{AII}$ & Symplectic & $-1$ & $0$ & $0$ \\
$\mathrm{CII}$ & Chiral Symplectic & $-1$ & $-1$ & $1$ \\
\br
\end{tabular}

\vspace{0.25cm}

\begin{tabular}{|c|c|c|c|c|c|c|}
\br
\textbf{Class} & $\mathbf{d=1}$ & {\bf QS} & $\mathbf{d=2}$ & {\bf QS} &  $\mathbf{d=3}$ & {\bf QS}  \\
\mr
$\mathrm{A}$ & 0 & & $\mathbb{Z}$& $\mathrm{yes}$ & $0$ & \\
$\mathrm{AIII}$ &  $\mathbb{Z}$ & yes & $0$ & & $\mathbb{Z}$& yes \\
$\mathrm{AI}$  & $0$ & & $0$ & & $0$ & \\
$\mathrm{BDI}$ & $\mathbb{Z}_2$&?&$\mathbb{Z}$&? & $0$& \\
$\mathrm{AII}$ & $0$&&$\mathbb{Z}_2$&yes & $\mathbb{Z}_2$&yes \\
$\mathrm{CII}$ & $2 \mathbb{Z}$&yes &$0$& & $\mathbb{Z}_2$&yes \\
\br
\end{tabular}
\end{center}

\end{table}

\subsubsection*{Bottom-Up Approach.} 

In this case, one  designs the ultracold-atom  Hamiltonian  so that it simulates  a particular model  belonging to the desired class of topological insulators.  Therefore, a  different experiment would be required for each class-oriented simulator. Two representative, yet reasonably simple examples are the Su-Schrieffer-Hegger model of polyacetilene~\cite{polyacetilene}, which is related to the $D=1$ BDI topological insulator, or the $\pi-$flux phase of the fermionic Creutz ladder~\cite{creutz_ladder},  which is related to  the $D=1$ AIII topological insulator. The former can be simulated by using a one-component Fermi gas in a one-dimensional dimerized optical superlattice, thus obtaining
\begin{equation}
\nonumber
H_{\mathrm{BDI}}=\sum_n (t-\delta)c^{\dagger}_{2n-1}c_{2n}+(t+\delta)c^{\dagger}_{2n}c_{2n-1} + \mathrm{H.c.},
\end{equation}
where $\delta$ quantifies the different tunneling strength between superlattice sites~\cite{PachosIgnacio}. On the other hand, the Creutz ladder is described by 
\begin{eqnarray}
\nonumber
H_{\mathrm{AIII}}=&\sum_{n}K\ee^{-\ii\theta}a^{\dagger}_{n+1}a_n+K\ee^{\ii\theta}b^{\dagger}_{n+1}b_n +Kb^{\dagger}_{n+1}a_n\\
&+Ka^{\dagger}_{n+1}b_n+M a^{\dagger}_{n}b_n+\mathrm{H.c.},
\end{eqnarray}
where $K,M$ are tunneling strengths, and $\theta$ is a magnetic flux piercing the ladder. This requires  
two Zeeman sublevels to be assigned to the fermion species $a_n,b_n$, and a one-dimensional laser-assisted tunneling $U_{\mathbf a_1}=\mathrm{diag}\{\ee^{-\ii\theta},\ee^{\ii\theta}\}$, $\tilde{U}_{\mathbf a_1}=\ii\sigma_x$, together with the Raman on-site operator of strength $M$~\cite{creutz_atoms}. 

It is possible to continue this approach, proceeding thus to higher dimensions and different topological classes. Prominent examples would be the honeycomb time-reversal breaking Haldane model~\cite{haldane,haldane_cold_atoms} for the $D=2$ topological insulator in class A, or the time-reversal Kane-Mele model in the honeycomb lattice in class AII~\cite{kane_mele}, or other optical-lattice geometries~\cite{SpielmanTI, kane_mele_atoms, bercioux}. Rather than following this route, we shall explore a different approach that is better suited to the superlattice-based simulator introduced above. Indeed, we shall argue that this quantum simulator allows the reproduction of most the topological phases in table~\ref{ti_table}.

\subsubsection*{Dimensional-Reduction Approach.} 

In this case, the starting point is the quantum simulator of $D$-dimensional Kaplan fermions  in  equation~(\ref{wilson_mass}). Depending on the particular choice of Dirac matrices,  the inverted-mass regime shall correspond to a  different class of  topological insulators. Besides,  in some situations, a dimensional reduction~\cite{dim_reduction} that amounts to the increase of the optical-lattice depth in one direction,  connect us to a different lower-dimensional class. We rewrite the full Hamiltonian 
\begin{equation}
 \label{kaplan}
H_{\mathrm{eff}}=\sum_{ \mathbf d, \mathbf{p}_{\mathbf{d}}}\Psi^{\dagger}( \mathbf{p}_{\mathbf{d}}) H_{\mathrm{DI}}^{\mathbf{d}}\Psi( \mathbf{p}_{\mathbf{d}}),\hspace{1ex}   H_{\mathrm{DI}}^{\mathbf{d}}(\mathbf{p}_{\mathbf{d}})=c\mathbf{\alpha}^{\mathbf{d}}\cdot\mathbf{p}_{\mathbf{d}}+m_{\mathbf{K}_{\mathbf{d}}}c^2\beta,
 \end{equation}
where the Wilson mass is $m_{\mathbf{K}_{\mathbf{d}}}=m-\sum_{\nu}(-1)^{d_{\nu}}m_{\nu}$, and where the Dirac matrices  $\alpha^{\mathbf{d}}_{\nu},\beta$ shall be selected so that the $\mathcal{T},\mathcal{C},\mathcal{S}$ symmetries are explicitly broken or preserved. This translationally-invariant Hamiltonian preserves these symmetries when the following conditions are met
\begin{eqnarray}
\mathcal{T} :&\qquad& \Theta^{\dagger}_{\mathrm{T}} \, \left[ H^{{\mathbf{d}}}_{\mathrm{DI}}(-\mathbf{p}_{\mathbf{d}}) \right]^* \, \Theta_{\mathrm{T}} \, = \, +H^{\mathbf{d}}_{\mathrm{DI}}(\mathbf{p}_{\mathbf{d}}); \\
\mathcal{C} :&\qquad&  \Theta^{\dagger}_{\mathrm{C}} \, \left[ H^{{\mathbf{d}}}_{\mathrm{DI}}(-\mathbf{p}_{\mathbf{d}}) \right]^* \, \Theta_{\mathrm{C}} \, = \, -H^{\mathbf{d}}_{\mathrm{DI}}(\mathbf{p}_{\mathbf{d}}); \\
\mathcal{S} :&\qquad&  \left[\Theta^{\dagger}_{\mathrm{T}}\right]^* \, \Theta^{\dagger}_{\mathrm{C}} \, H^{{\mathbf{d}}}_{\mathrm{DI}}(\mathbf{p}_{\mathbf{d}}) \, \Theta_{\mathrm{C}}\, \Theta_{\mathrm{T}}^* \,=\,-H^{\mathbf{d}}_{\mathrm{DI}}(\mathbf{p}_{\mathbf{d}});
\end{eqnarray}
where $\Theta_{\mathrm{T}},\Theta_{\mathrm{C}}$ are some unitary matrices. In table~\ref{ti_table_bis}, we list the symmetry properties of different Kaplan-fermion Hamiltonians. It is important to note that these symmetries might correspond to the exact symmetries in nature (e.g. when considering the hyperfine levels $\{\ket{F,m_F},\ket{F,-m_F}\}$, the time-reversal symmetry given by $\theta_{\mathrm{T}}=\ii\sigma_y$ exactly correspond to time-reversal symmetry in nature). Conversely, these symmetries might otherwise correspond to the algebraic properties of the effective Hamiltonian. Let us emphasize, however, that as far as the disorder respects such symmetries, the robustness of the edge excitations is guaranteed. It would be of the greatest interest to design disorder breaking or preserving such symmetries, generalizing the studies on Anderson localization~\cite{Aspect, anderson_atoms}.

\begin{table}
\caption{ Quantum simulator of Topological Insulators. We list different realizations of Wilson-fermion Hamiltonians in equation~(\ref{kaplan}) that directly lead to several classes of topological insulators. Each class, characterized by the discrete  symmetries $\mathcal{T},\mathcal{C},\mathcal{S}$, where $\mathcal{T}^2=\Theta_{\mathrm{T}}\Theta_{\mathrm{T}}^*=\pm 1$, and $\mathcal{C}^2=\Theta_{\mathrm{C}}\Theta_{\mathrm{C}}^*=\pm 1$. Besides, each class has a Wilson-fermion representative with a particular choice of the Clifford algebra $\alpha_{\nu},\beta$ that depends on the dimensionality $D$ and the corresponding symmetries. We also highlight the topological insulators that can be obtained by dimensional reduction from a parent Hamiltonian, such as $\mathrm{AII}, D=3\hookrightarrow \mathrm{AII}, D=2$, or $\mathrm{A}, D=2\hookrightarrow \mathrm{AIII}, D=1$. We also list the unitary matrices $\Theta_{\mathrm{T}}, \Theta_{\mathrm{C}}$ involved in the definition of the discrete symmetries. }
  \label{ti_table_bis}
\begin{center}
\begin{tabular}{| l | c | c | c | c | c | c | c | c | c | c | c|}
\br
\textbf{Class} & $\mathbf{D}$ & $\bm{\alpha_x}$ & $\boldsymbol{\alpha_y}$ & $\boldsymbol{\alpha_z}$ & $\boldsymbol{\beta}$ & $\boldsymbol{\Theta}_{\mathrm{T}}$ & $\boldsymbol{\Theta}_{\mathrm{C}}$ & $\mathcal{T}$ & $\mathcal{C}$ & $\mathcal{S}$ \\
\mr
CII & $3$  & $\sigma_z\otimes\sigma_x$ & $\sigma_z\otimes\sigma_y$ & $\sigma_z\otimes\sigma_z$ & $\sigma_z\otimes\mathbb{I}_2$ & $\ii\mathbb{I}\otimes\sigma_y$ & $\ii \sigma_x\otimes\sigma_y$ & -1 & -1 & 1 \\
\mr
AIII & $3$ & $\sigma_z\otimes\sigma_x$ & $\sigma_z\otimes\sigma_y$ & $\sigma_z\otimes\sigma_z$ & $\sigma_y\otimes\mathbb{I}_2$ & $\ii\mathbb{I}\otimes\sigma_y$ & $\ii \sigma_x\otimes\sigma_y$ & 0 & 0 & 1 \\
\mr
AII & $3$  & $\sigma_z\otimes\sigma_x$ & $\sigma_z\otimes\sigma_y$ & $\sigma_z\otimes\sigma_z$ & $\sigma_x\otimes\mathbb{I}_2$ & $\ii\mathbb{I}\otimes\sigma_y$ & $\ii \sigma_x\otimes\sigma_y$ & -1 & 0 & 0  \\
\mr
$\hookrightarrow$ AII & $2$  & $\sigma_z\otimes\sigma_x$ & $\sigma_z\otimes\sigma_y$  & &  $\sigma_x\otimes\mathbb{I}_2$ & $\ii\mathbb{I}\otimes\sigma_y$ & $\ii \sigma_x\otimes\sigma_y$ & -1 & 0 & 0 \\
\mr
A & $2$ & $\sigma_x$ & $\sigma_y$ &  & $\sigma_z$ & $\mathbb{I}$ & $\ii \sigma_x$ & 0 & 0 & 0 \\
\mr
$\hookrightarrow$ AIII & $1$  &  $\sigma_x$ &&&$\sigma_z$ &  $\mathbb{I}$ & $\ii \sigma_x$ & 0 & 0 & 1\\
\mr
CII & $1$ & $\sigma_z\otimes\sigma_x$ & & & $\sigma_z\otimes\mathbb{I}_2$ & $\ii\mathbb{I}\otimes\sigma_y$ & $\ii \sigma_x\otimes\sigma_y$ & -1 & -1 & 1 \\
\br
\end{tabular}
\end{center}

\end{table}

In table~\ref{ti_table_bis}, we have listed the different topological insulators that can be simulated with our scheme. As shown in~\cite{MazzaRizziPRL} for the particular case of three-dimensional
AII insulators, the laser parameters can be controlled so that an {\it odd} number of Wilson masses are inverted. This mass-inversion occurs through a gap-closing point, and thus a quantum
phase transition between a normal band insulator, and a topological one occurs. This new phase is characterized by an $odd$ number of massless fermionic excitations (i.e. massless Dirac fermions) bound to the boundaries of  the system, and protected by a topological invariant. In the three-dimensional case, this corresponds to an axion term that modifies the response of the system according to the so-called axion electrodynamics~\cite{axion}. Remarkably, table~\ref{ti_table_bis} contains all the relevant information to explore the exotic properties of different topological insulators in a superlattice based experiment with ultracold atoms.

\section{Conclusions \label{sec:conclusion}}

In this article, we have presented a concrete proposal for the realization of laser-assisted tunneling in a spin-independent optical lattice trapping a multi-spin atomic gas. Remarkably enough,  it is possible to tailor a wide range of spin-flipping hopping operators, which opens an interesting route to  push the experiments beyond the standard superfluid - Mott insulator transition. The scheme we have presented combines bi-chromatic lattices and Raman transfers, to adiabatically eliminate the states trapped in the middle of each lattice link. These states act as simple spectators that allow us to assist the tunneling of atoms between the main minima of the optical lattice. This mechanism is clearly supported by our numerical simulations of the time-evolution of the atomic population between the different optical-lattice sites. Even if we focus on fermionic $\pot$K, we stress that  the ingredients of this proposal do not rely on the atomic statistics, and could be thus used  for all the alkalis.

We believe that such a device could have important applications in the quantum simulation of non-interacting lattice field theories, which are characterized, in their discrete version, by on-site and nearest-neighbor hopping Hamiltonians.
Once the fields of the theory to be simulated are mapped into the atomic hyperfine states, the desired operators correspond to population transfers between such levels. The former can be realized by standard microwaves, whereas the latter might be tailored with the laser-assisted schemes  described here.

Even though interactions are at the heart of a plethora of interesting effects, non-interacting fermionic theories already encompass a number of phenomena whose experimental realization would be of the greatest interest.
In the second part of the article, we  analyzed interesting physical models which could benefit from our proposal. In particular, we focused on  relativistic field theories, and showed that there is a zoo of relativistic lattice fermions that can be addressed with this platform. Besides, we presented a toolbox to design particular assisted tunneling processes that lead us to the physics of topological insulators. Remarkably enough, this quantum simulator turns out to be extremely versatile, since most of the phases of the periodic table of topological insulators can be addressed. 

Finally, let us comment on the possible combination of this proposal with the control of interactions already achieved in cold-atom gases. This might eventually boost experiments into regimes where classical numerical simulations fail, which we leave here as an outlook for future work.
In particular, the problem of robustness of topological orders (classified for non-interacting theories) with respect to interactions is one of the most important challenges of the modern condensed matter~\cite{inter:top}.
We believe that a direct combination of our setup with Feshbach resonances will provide important insights into this unsolved question.

\ack
LM and MR warmly acknowledge fruitful discussions with U. Schneider, with whom the superlattice idea was actually conceived. The authors also thank J.I. Cirac, G. Juzeliunas, W. Phillips, I. Spielman, and C. Wu.
LM and MR thank Caixa Manresa and ICFO for hospitality; MR and ML thank KITP for hospitality.
MR and ML have received funding from the European Community's Seventh Framework Programme (FP7/2007-2013) under grant agreement no. 247687 (IP-AQUTE).
ML also acknowledges financial support from ERC Grant QUAGATUA, EU STREP NAMEQUAM, MINCIN FIS2008-00784, Alexander von Humbold Foundation and Hamburg Theory Award.
NG thanks the FRS-FNRS for financial support.
AB and MAMD thank MICINN FIS2009-10061, CAM QUITEMAD, European FET-7 PICC, UCM-BS GICC-910758.

\end{document}